\documentclass[sigconf]{acmart}
\AtBeginDocument{%
  }

\copyrightyear{2025} 
\acmYear{2025} 
\setcopyright{acmlicensed}\acmConference[IUI '25]{30th International Conference on Intelligent User Interfaces}{March 24--27, 2025}{Cagliari, Italy}
\acmBooktitle{30th International Conference on Intelligent User Interfaces (IUI '25), March 24--27, 2025, Cagliari, Italy}
\acmDOI{10.1145/3708359.3712162}
\acmISBN{979-8-4007-1306-4/25/03}

\usepackage{multirow}

\begin{document}

\title{An Exploratory Study on How AI Awareness Impacts Human-AI Design Collaboration}


\author{Zhuoyi Cheng}
\orcid{0009-0003-3941-8869}
\affiliation{%
  \department{College of Software Technology}
  \institution{Zhejiang University}
  \city{Hangzhou}
  \country{China}
}
\email{zy_cheng@zju.edu.cn}

\author{Pei Chen}
\orcid{0000-0003-0962-6459}
\authornote{Corresponding author}
\affiliation{%
  \department{College of Computer Science and Technology}
  \institution{Zhejiang University}
  \city{Hangzhou}
  \country{China}
}
\email{chenpei@zju.edu.cn}

\author{Wenzheng Song}
\orcid{0009-0003-1271-7623}
\affiliation{%
  \department{College of Computer Science and Technology}
  \institution{Zhejiang University}
  \city{Hangzhou}
  \country{China}
}
\email{svvz@zju.edu.cn}

\author{Hongbo Zhang}
\orcid{0000-0001-5376-9460}
\affiliation{%
  \department{College of Computer Science and Technology}
  \institution{Zhejiang University}
  \city{Hangzhou}
  \country{China}
}
\email{hongbozhang@zju.edu.cn}

\author{Zhuoshu Li}
\orcid{0009-0000-3327-1622}
\affiliation{%
  \department{College of Computer Science and Technology}
  \institution{Zhejiang University}
  \city{Hangzhou}
  \country{China}
}
\email{lizhuoshu@zju.edu.cn}

\author{Lingyun Sun}
\orcid{0000-0002-5561-0493}
\affiliation{%
  \department{College of Computer Science and Technology}
  \institution{Zhejiang University}
  \institution{Zhejiang – Singapore Innovation and AI Joint Research}
  \city{Hangzhou}
  \country{China}
}
\email{sunly@zju.edu.cn}

\renewcommand{\shortauthors}{Cheng et al.}

\begin{abstract}
The collaborative design process is intrinsically complicated and dynamic, and researchers have long been exploring how to enhance efficiency in this process. As Artificial Intelligence technology evolves, it has been widely used as a design tool and exhibited the potential as a design collaborator. Nevertheless, problems concerning how designers should communicate with AI in collaborative design remain unsolved. To address this research gap, we referred to how designers communicate fluently in human-human design collaboration, and found awareness to be an important ability for facilitating communication by understanding their collaborators and current situation. However, previous research mainly studied and supported human awareness, the possible impact AI awareness would bring to the human-AI collaborative design process, and the way to realize AI awareness remain unknown. In this study, we explored how AI awareness will impact human-AI collaboration through a Wizard-of-Oz experiment. Both quantitative and qualitative results supported that enabling AI to have awareness can enhance the communication fluidity between human and AI, thus enhancing collaboration efficiency. We further discussed the results and concluded design implications for future human-AI collaborative design systems.
\end{abstract}




\maketitle

\section{Introduction}
The design process is intrinsically improvisational, open-ended, and dynamic~\cite{Rezwana2023COFI, Zhou2024Understanding}, in which designers occasionally switch between different design phases and iterate their ideas to achieve an optimal design~\cite{Hybs1992an}. To enhance efficiency in such a complicated process, researchers have long been studying how to integrate Artificial Intelligence~(AI) into the design process~\cite{Eckert2000Intelligent, Davis2016Empirically, Wang2024RoomDreaming}. Thanks to the rapid evolution in large-scale models, current AI technologies have gained higher creativity, stronger reasoning~\cite{Wang2024ModaVerse, Girdhar2023ImageBind}, and generation abilities~\cite{Shen2024PMG, Edwards2024Sketch2Prototype}, making it increasingly suitable for enhancing the efficiency of the design process. Prior endeavors have applied AI in different design stages, like ideation~\cite{Koch2019May, Xu2024Jamplate}, prototyping~\cite{Malsattar2019Designing, Dering2017an}, and evaluation~\cite{Camburn2020Computer, Zhang2017Computational}. In these cases, AI was treated as a tool to automate certain design tasks, and most interactions were initiated by designers. However, such AI tools may introduce new interaction paradigms and bring extra work in steering the system to obtain ideal feedback, which may increase designers' workload and learning cost~\cite{Gmeiner2023Exploring}. 

More recently, researchers have sought to enable AI to directly collaborate with designers. For example, \citet{muller2024group} designed an AI chatbot capable of autonomously suggesting ideas in the group brainstorming process, and \citet{Kuang2024UX} engaged UX designers to inspect UX problems with AI collaboratively. In the general area of human-AI collaboration, researchers have exploited different theories and frameworks to model human-AI collaboration, facilitating the understanding of human-AI collaboration at an abstract level. For instance, \citet{Tankelevitch2024Metacognitive} applied the descriptive framework for metacognition to understand users' metacognitive demands when using Generative AI systems, and provided cases to lower the demands. \citet{Muller2022Extending} described human-AI collaboration pattern by illustrating how human and AI initiative shifts in different scenarios. However, communication between human and AI remains to be explored extensively, which is crucial for decision-making, design information sharing, and shared understanding building in the highly dynamic design process~\cite{Chiu2002187an, Tan2023performance, Krishnakumar2022story},

To address this research gap, we first looked into existing studies about human-human collaboration to understand how designers communicate with collaborators fluently~\cite{Wardak2016Gestures, Baird2000ethnographic, Nelson2020Opening}. In various forms of human-human design collaboration, including face-to-face~\cite{Wardak2016Gestures}, virtual~\cite{Koutsabasis2012on}, and online~\cite{Vyas2013studio} collaborations, designers comprehend the status quo and leverage information in the environment and communicative cues from collaborators to initiate verbal communication. Having awareness of collaborators and the current situation is the key to such processes~\cite{Gutwin2002descriptive, Endsley2023Supporting, Vyas2013studio}. There are various definitions for the term ``awareness'' (e.g.,~\cite{Endsley2023Supporting, Gutwin2002descriptive, Gutwin2004Group}), but most of them share the core meaning concluded by~\citet[p.~107]{Dourish1992awareness}, which we used in this study: ``\textit{...awareness is an understanding of the activities of others, which provides a context for your own activity.}'' The awareness activity happens naturally in co-located collaboration, while awareness support is needed if people are working distributedly in a collaborative system~\cite{Gutwin2002descriptive}. For instance, \citet{Cidota2016HMD} evaluated different notification systems for users virtually collaborating in an augmented reality (AR) environment to develop users' awareness of each other. \citet{Dehler2011guiding} developed awareness support in a collaborative learning environment to assist learners in having awareness of others' knowledge. We propose that quipping AI with this ability can assist in efficient human-AI communication and collaboration. However, while humans can naturally be aware of others and the environment~\cite{Gutwin2002descriptive}, enabling AI to have awareness of humans is underexplored in both the impacts of doing so and the way to do so.

To this point, we explored the impacts of enabling AI awareness in human-AI design collaboration, and our main research question (RQ) focused on ``\textbf{How will AI having awareness of the designer and current situation impact communication in human-AI design collaboration?}'' As the RQ pertains to a future system that does not exist yet, we conducted user studies using the Wizard-of-Oz (WoZ) method. We first implemented a system that ensured the basic design and human-AI collaboration functions. Next, we iteratively designed a WoZ workflow with two Wizards to complement the communication and awareness functions. We then conducted a within-subject experiment involving 20 participants using this system and extracted quantitative and qualitative data from the experiment process recordings, conversation histories, and interview transcripts. The quantitative results showed that human-AI communication was significantly more fluent when AI had awareness, implying more efficient collaboration, and thematic analysis results of the interview presented detailed impacts on communication in human-AI collaboration. We further discussed the above results to build relationships between quantitative and qualitative data to reveal the causalities and inferences. Last, we concluded design implications about maintaining and leveraging awareness in human-AI collaboration and communication strategies for AI in future human-AI collaborative design systems.

Our research provides two main contributions to future human-AI collaborative design. First, we proposed a prospecting way to enhance communication fluidity in human-AI design collaboration by enabling AI to have awareness of designers and the current situation, and conducted user studies to certify this prospect. Second, we provided novel insights into future development of human-AI collaborative design systems based on the experiment results. We believe this study can shed light on how human and AI should communicate and collaborate not only in the design process, but also in other processes that share the complex and dynamic features with the design process.

\section{Related Work}
To better situate this study and provide references for system design, we conducted a thorough literature review of research on the following three topics: 1) how do researchers integrate AI in the design process, including AI-driven design tools and AI collaborators, 2) common communication channels in collaboration and how they were transferred to human-AI collaboration, and 3) mechanisms for supporting awareness in co-located and distributed human-human collaboration, and human-AI collaboration.

\subsection{Integrating AI in Design}
Design requires knowledge, information, and skills from diverse areas, and has long been a task that demands efforts beyond a single designer due to its burgeoning complexity~\cite{Rodgers1998role, Tang1996AI, Zhou2024Understanding}. The need for collaboration and assistance thus naturally appeared, and AI is one important technology that can help. In early research, AI was considered as a design support tool to assist in quantifiable or functional design tasks~\cite{Rodgers1998role}. For example, \citet{Chakrabarti1994Two} proposed a two-step approach to work out viable mechanical design solutions based on a set of known rules, and \citet{Grecu1996Learning} explored AI support for parametric design, where the product has already been defined except for combining several detailed values (e.g., color) with various options. In these tasks, AI was expected to reduce tedious work and give predictable outputs. As research on computational creativity booms, AI gained creativity and moved from the above ``routine design tasks'' to ``creative design tasks'' in the design process. Designers can now use various AI-driven tools to obtain ideas~\cite{Liu20233DALL, Koch2019May}, organize thoughts~\cite{Zhong2024Causal}, explore design space~\cite{Camburn2020Computer, Wang2024RoomDreaming}, etc. However, these applications are mainly design tools that require designers to initiate operations.

Recently, the development of AI abilities in generation and reasoning abilities~\cite{Wang2024ModaVerse, Edwards2024Sketch2Prototype} allowed AI to take an active role in design~\cite{Li2023Impression}, thus being prospective in becoming a design collaborator. Researchers are striving for better human-AI design collaboration from both theoretical and practical perspectives. \citet{Rezwana2023COFI} did a thorough literature review on human-AI co-creation and proposed a COFI framework to describe the interactions between collaborators and with the shared product in co-creative systems. \citet{Shi2023Understanding} also reviewed articles but specifically focused on designer-AI collaboration, and concluded five characteristics, including scope, access, agency, flexibility, and visibility, to help describe designer-AI collaboration and identify existing problems. Results from both works elaborated on problems in human-AI communication, with the former one presenting the lack of ``human to AI consequential communication'' and ``AI to human communication'', and the latter one pointing out the importance and a dearth of multiple communication modalities other than visual ones. Other more practice-oriented studies focus on specific design tasks, exploring interaction pathways and guidelines for human-AI collaboration. For example, \citet{Zhou2024Understanding} explored the nonlinear collaboration between graphic designers and AI in the OptiMuse system using the WoZ method, and proposed a co-design framework. \citet{Pan2023Diagrams} explored how AI can intelligently collaborate with people through a WoZ-like method, and implemented a collaborative diagram editing system guided by obtained insights.

Similar to \citet{Zhou2024Understanding} and \citet{Pan2023Diagrams}, our work seeks to explore the human-AI collaborative design process. Grounded in the problems identified by \citet{Rezwana2023COFI} and \citet{Shi2023Understanding}, we specifically focus on problems in human-AI communication and explore the potential of AI awareness to alleviate this problem.

\subsection{Communicating in Human-Human and Human-AI Collaboration}
In traditional human-human co-located design collaboration, people share expertise, ideas, resources, or responsibilities in the design process, to which communication is crucial~\cite{Chiu2002187an}. Verbal communication is the most common way that people exchange information. People leverage various linguistic strategies and non-verbal communicative cues to enhance the efficiency of verbal communication, for example, deictic words can specify the content of the current discussion in a highly contracted way~\cite{Tory2008deictic}, gestures can focus attention and represent ideas or features, and facial expressions can assist in judging one's true feelings~\cite{Ge2021emotion}. When collaboration is moved online and people work distributedly through collaborative systems, verbal communication, and its corresponding communication strategies require special support to function~\cite{Gutwin2002descriptive}. However, as remote collaboration becomes common and frequent, collaborative systems using text-based communication have matured rapidly~(e.g., Comment function in Figma\footnote{\url{https://www.figma.com/}} and Miro\footnote{\url{https://miro.com}}).

Similar to human-human distributed collaboration, communication and interaction in current human-AI collaboration is mainly based on text-based conversations. For example, \citet{muller2024group} implemented a text-based chatbot to facilitate the brainstorming process, and \citet{Kuang2024UX} designed a collaborative system for UX designers to identify usability problems with text-based conversation AI assistants. However, communicating with AI through text is not necessarily an optimal choice for human-AI communication, especially in the design process. Previous research has found communicating through voice leads to higher trust in both humans and chatbots compared to text~\cite{Burri2018, Bente2008Networking}. Additionally, the need for composing textual prompts and conducting other operations through visual channel can cause increased cognitive load~\cite{Tankelevitch2024Metacognitive, Vermeulen2008Sensory}, thus making it harder for mutual understanding in the design process~\cite{Gmeiner2023Exploring}. In the more general field of human-AI collaboration, researchers have found verbal communication to be a favored communication strategy because of its direct and efficient feature in difficult tasks~\cite{Zhang2021ideal}. In a recent research about sketching, researchers pointed out that people ``\textit{usually use language to talk about real, distant, invisible, imaginary, or conceptual objects}''~\cite[p.~4-5]{Rosenberg2024DrawTalking}, which are similar to what happens in the design process~\cite{Kokotovich2016abstraction, Ungureanu2021Analysing}. However, systems supporting designers to communicate with AI through speech are underexplored~\cite{Rezwana2023COFI, Shi2023Understanding}. Therefore, we allow designers to communicate with AI through speech in this study to gain insights for future human-AI collaborative design systems that support verbal communication.

\begin{figure*}[!th]
    \centering
    \includegraphics[width=1\linewidth]{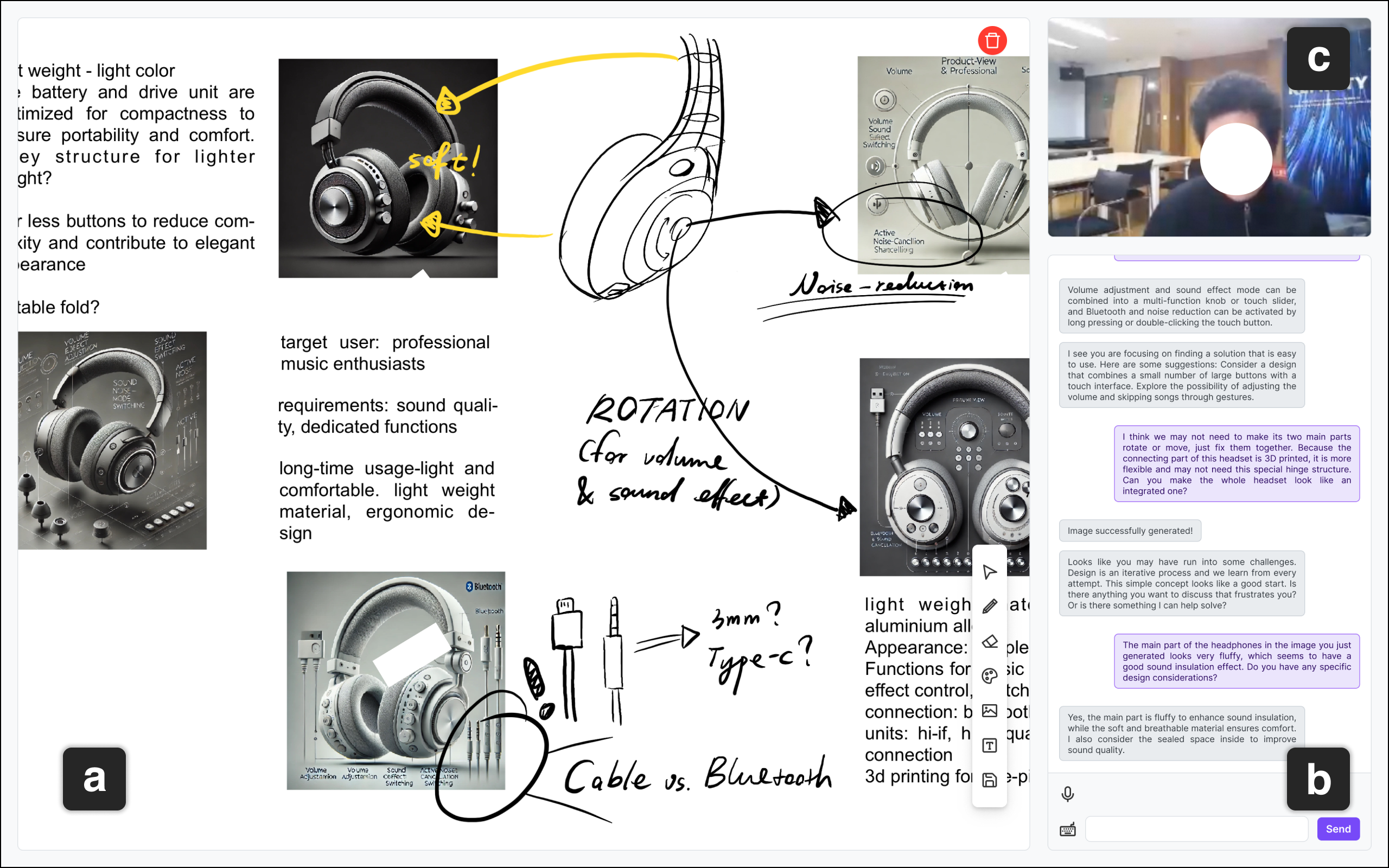}
    \caption{The interface example of the human-AI design collaborative system: a) Canvas. Canvas is the main interaction area where designers edit information using tools in the toolbar on the bottom right corner, and AI can also upload generated images to the canvas. b) Chat box. Communication history and speech recognition results are displayed in this area, and designers can also type here to initiate conversation in case of need. c) Camera screen. This is presented \textit{in the Aware condition only} to indicate what AI can see. \textit{Note: The interface is a translated version of the original one used for user studies.}}
    \label{fig:interface}
    \Description{This figure illustrates the system interface. The interface consists of three areas: canvas, chat box, and camera screen. The canvas, marked as area A, is on the right side, taking up about three-quarters of the interface. There is a user case from one participant, a toolbar in the bottom right corner, and a delete button in the top right corner of the canvas. The chat box and camera screen areas together take up about a quarter of the interface, marked as area B and area C respectively. The chat box takes up more than two-thirds of this area, with some communication histories of the participant and AI presenting in it and an input area for presenting speech recognition results at the bottom. The camera screen area takes up less than a third of this area, displaying the captured screen.}
\end{figure*}

\subsection{Supporting Awareness in Collaboration}
The term ``awareness'' had been combined with different words to coin a new term and specify a specific research scope, like shared awareness~\cite{Chakrabarti1994Two} and situation awareness~\cite{Endsley1988Design}. Similar as their overall meanings were, ``awareness'' has not been defined and used consistently~\cite{Gross2013Supporting}. In this paper, we adopt the definition provided by~\citet[p.~107]{Dourish1992awareness}:

\begin{quote}
    \textit{``[...]awareness is an understanding of the activities of others, which provides a context for your own activity. This context is used to ensure that individual contributions are relevant to the group's activity as a whole, and to evaluate individual actions with respect to group goals and progress. The information, then, allows groups to manage the process of collaborative working.''}
\end{quote}

Awareness is critical to successful design collaboration for providing information including shared information, group and individual activities, and coordination, to contextualize one's activity~\cite{Dourish1992awareness}. In co-located human-human collaboration, awareness activities are supported by human perceptions, which can happen naturally~\cite{Gutwin2002descriptive}. However, when collaboration is distributed, awareness activities require supporting mechanisms in the collaborative system to proceed~\cite{Gutwin2002descriptive}, and researchers have developed various mechanisms to enable people to have awareness of other collaborators and the environment in different scenarios. \citet{Gutwin1996Widgets} designed a set of interface widgets to support distributed workers having awareness of collaborators' current activities and view, including radar view, WYSIWIS (What You See Is What I See) view, WYSIWID (What You See Is What I Do) view, etc. \citet{Cidota2016HMD} compared different notification mechanisms for people collaborating through AR. \citet{Dehler2011guiding} designed a collaborative system and presented participants' knowledge levels to allow a better understanding of the group and each other. \citet{Janssen2011Group} visualized group members' contributions in online collaboration to examine how collaborative learning is impacted by group awareness.

Although studies and applications of awareness mechanisms in human-human collaboration are abundant, the idea of equipping AI with awareness ability is underexplored due to the infancy of human-AI collaboration research and the difficulty of transferring this human ability to AI. However, previous research has demonstrated that allowing users to have awareness of AI presence positively contributes to user engagement and collaborative experience~\cite{Rezwana2021penpal}, implying the potential advantages of applying the awareness theory in human-AI collaboration. Therefore, in this study, we first explored the impact of allowing AI to have awareness (i.e., AI is aware of the designer's design activities and current working content) through the WoZ method to ground future design and research. By exploring the potential of transferring the ``awareness'' concept in human-AI collaboration, we expect to guide the design of AI collaborators that have a certain level of initiative in a shared workspace~\cite{Salikutluk2024situational, He2024Canvas}, adjust their output according to real-time contextual information~\cite{Ma2023ProactiveAgent}, and infer deeper reasons and meaning behind the literal senses of obtained visual and textual information~\cite{Liu2024ComPeer, Liang2019Implicit} to support more dynamic and difficult tasks like design through human-AI collaboration.

\begin{figure*}[bthp]
    \centering
    \includegraphics[width=1\linewidth]{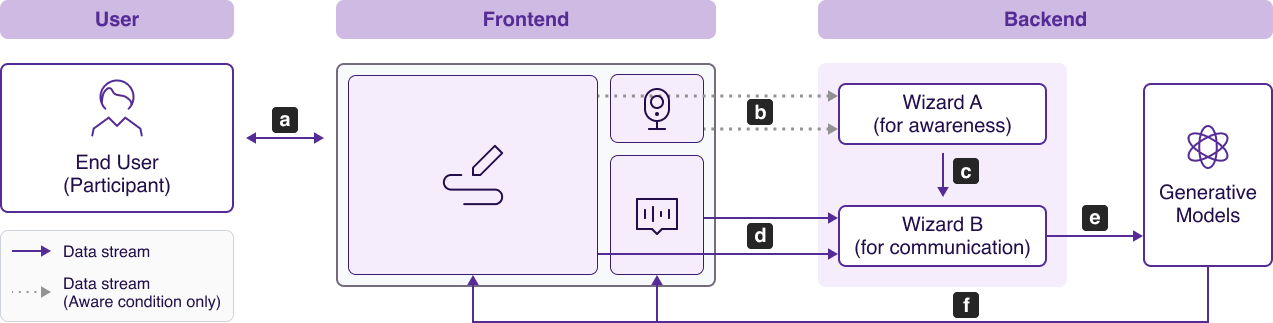}
    \caption{The overview of system structure and data stream. In our study, the participants \textbf{a)} interacted and communicated with the frontend (i.e., the user interface in Figure~\ref{fig:interface}) and received feedback from AI. The backend consisted of two Wizards and generative models. In the Aware condition, Wizard A would \textbf{b)} obtain awareness information from the camera screen and canvas, \textbf{c)} compose prompts based on predefined scripts and send them to Wizard B. In the Non-aware condition, the prompt sent in process \textbf{(c)} would not contain awareness information. Wizard B in both conditions supported the human-AI communication process by \textbf{d)} processing speech recognition results and input with canvas information, and \textbf{e)} sending prompts coming from the participants and Wizard A to generative models. \textbf{f)} All generation results would be uploaded to the canvas.}
    \Description{The overview of system structure and data stream. Areas from left to right are named: User area, Frontend area, Backend area. The user area presents a box representing the end user, the frontend area presents a draft of the system interface described in Figure 1, and the backend area consists of three boxes, with two boxes symbolizing two wizards in respective, and the remaining box symbolizing the generative models at the backend. The data streams are illustrated by two types of arrows, with the solid purple arrows representing data streams in both the Aware and Non-aware conditions, and the dashed gray arrows only representing data streams in the Aware condition. Data stream A describes the mutual data exchange between the user and the frontend, illustrated by a solid purple double arrow connecting the user area and the frontend area. Data stream B describes Wizard A's comprehension of awareness information, gained from the canvas and the camera screen, and is illustrated by two dashed gray arrows, pointing from the canvas and camera screen areas in the frontend area to Wizard A in the backend area. Data stream C describes the process of Wizard A sending prompts to Wizard B, illustrated by a solid purple arrow pointed from Wizard A to Wizard B. Data stream D describes the process Wizard B receives information from the frontend, illustrated by two solid purple arrows pointed from the canvas and chat box areas in the frontend area to Wizard B in the backend area. Data stream E describes the process of Wizard B sending crafted prompts to the generative models, illustrated by a solid purple arrow pointed from Wizard B to Generative Models. Data stream F describes the process of generative models sending the outputs to the frontend, illustrated by two solid purple arrows pointed from the generative model in the backend area to the canvas and chat box areas in the frontend area.}
    \label{fig:systemStructure}
\end{figure*}

\section{Human-AI Collaborative Design System}
In this section, we introduced the system used in our study, including user interface, system functions, and implementations\footnote{For source code and detailed guidelines, please refer to the supplementary material or \url{https://github.com/SongWZ3214/AI_teammate/tree/master}.}.

\subsection{User Interface and Interactions}
As an initial prototype, we kept the interface simple and consistent with common human-AI collaboration systems~(e.g.,~\cite{Kuang2024UX, Zhou2024Understanding}), and only functions served the RQ will be added apart from basic design and collaboration functions.

The interface is implemented with the React frontend framework\footnote{\url{https://react.dev}} and Ant Design UI component\footnote{\url{https://ant.design/components/overview-cn/}}. The overall interface layout is consistent with current collaborative systems containing a shared workspace/canvas and a chatbot~\cite{Kuang2024UX, Reza2024ABScribe, Robe2022PairBuddy, Ross2023Programmer}. Three main areas consist of the interface (Figure~\ref{fig:interface}): a) Canvas. Basic design functions, including adding and removing text, images, and sketches are supported. Participants can switch between different tools in the toolbar to edit the canvas. b) Chat box. Participants can communicate with AI through speech or textual input, and we encouraged participants to communicate through speech in this study unless the information can not be conveyed with speech. When using speech input, participants can naturally initiate or terminate a conversation without saying activation or termination commands. All communication histories between the participant and AI will be preserved and displayed in this area. Besides, participants can communicate with AI using deictic words like ``this'' or ``its'' to indicate what they select, which is a common linguistic strategy to simplify communication~\cite{Tory2008deictic, Bolt1980Put, Rosenberg2024DrawTalking}. c) Camera screen. We use a Hikvision 2K Computer Camera to monitor the participants. They will not interact with this area, but can see what the camera is currently capturing, which equals what the AI is seeing. This area \textbf{\textit{only}} appears in the condition where AI has awareness (henceforth the Aware condition), and in the condition where AI does not have awareness (henceforth the Non-aware condition), this area is hidden.

\subsection{System and Wizard Implementation}
\label{wizardImplementation}
The system and Wizard implementation in the two conditions (Aware and Non-aware) are illustrated in Figure~\ref{fig:systemStructure}. All contents and activities that designers see and experience exist in the frontend (Figure~\hyperref[fig:systemStructure]{\ref*{fig:systemStructure}a}), which is supported by two Wizards and generative models at the backend (Figure~\hyperref[fig:systemStructure]{\ref*{fig:systemStructure}b-f}). The system backend is built with Flask\footnote{\url{https://flask.palletsprojects.com/en/3.0.x/}}, and its functions can be broken down into three parts: awareness mechanisms fulfilled by Wizard A, communication functions supported by Wizard B, and feedback generation completed by generative models.

\textit{Awareness mechanisms.} The awareness mechanisms will only function in the Aware condition (See Figure~\hyperref[fig:systemStructure]{\ref*{fig:systemStructure}b}). In this condition, Wizard A will observe and judge participants' current design activities per 60-90 seconds from participants' shared screen (displaying the frontend/system interface) through an online meeting software according to adapted coding scheme from~\cite{Kim2011Relations} (Table~\ref{tab:coding_scheme}) to minimize bias. This information is then filled in a prompt template, and together with the screenshot of the current working process on the canvas, will be sent to Wizard B through real-time messaging software for eliciting AI feedback. In the Non-aware condition, Wizard A will only send common prompts without awareness information or screenshots. The prompt templates used in the two conditions were designed and iterated in pilot studies, and are presented as follows:

\begin{quote}
    \textbf{The Aware Condition:}\\
    \texttt{We are currently in Step [ ], designer's design activity is [ ], and the content the designer is currently working on is presented in the image.\\
    Your feedback is:}
\end{quote}

\begin{quote}
    \textbf{The Non-aware Condition:}\\
    \texttt{We are currently in Step [ ], please think divergently/convergently according to Requirement [ ].\\
    Your feedback is:}
\end{quote}

\textit{Communication between designer and AI.} Wizard B is mainly responsible for organizing participants' input into prompts, receiving prompts from Wizard A, and sending these prompts to generative models. Participants in our study are instructed to mainly use speech communication. We used the Web Speech API\footnote{\url{https://developer.mozilla.org/en-US/docs/Web/API/Web_Speech_API}} for speech recognition (i.e., STT, Speech-to-Text) and Text-to-Speech (TTS) functions. All that the participants speak will be recognized and transcribed to text, and sent to the backend. Wizard B will first judge whether the participants intend to communicate with AI, if yes, Wizard A will then assemble the transcribed speech snippets at the backend console if a complete paragraph is fragmented into several parts to form a complete prompt. Undesired inputs (e.g., think aloud, murmurs, conversations from and with the experimenters) are ignored in this process. Prompts produced by Wizard B and come from Wizard A are all sent to generative models for feedback generation. Textual inputs sent by participants will be directly sent to the generative models by Wizard A. The above operations were processed at the backend using a series of predefined short-cut commands by Wizard B.

\textit{Feedback generation.} Feedback generation is supported by dedicated generative models for generating verbal/textual content and images in respective, which is GPT-4 turbo\footnote{\url{https://platform.openai.com/docs/models/gpt-4-and-gpt-4-turbo}} for verbal/textual content, and DALL·E 3\footnote{\url{https://openai.com/index/dall-e-3/}} for images. GPT-4 turbo will be initiated by the same prompt across participants (Appendix~\ref{initiate_prompt}) to ensure the models behave consistently across participants. The generated contents are then uploaded to the frontend. Images will be directly put on the canvas, verbal content will be converted into speech, and its corresponding text will be displayed in the chat box.

\begin{figure}[ht]
    \centering
    \includegraphics[width=\linewidth]{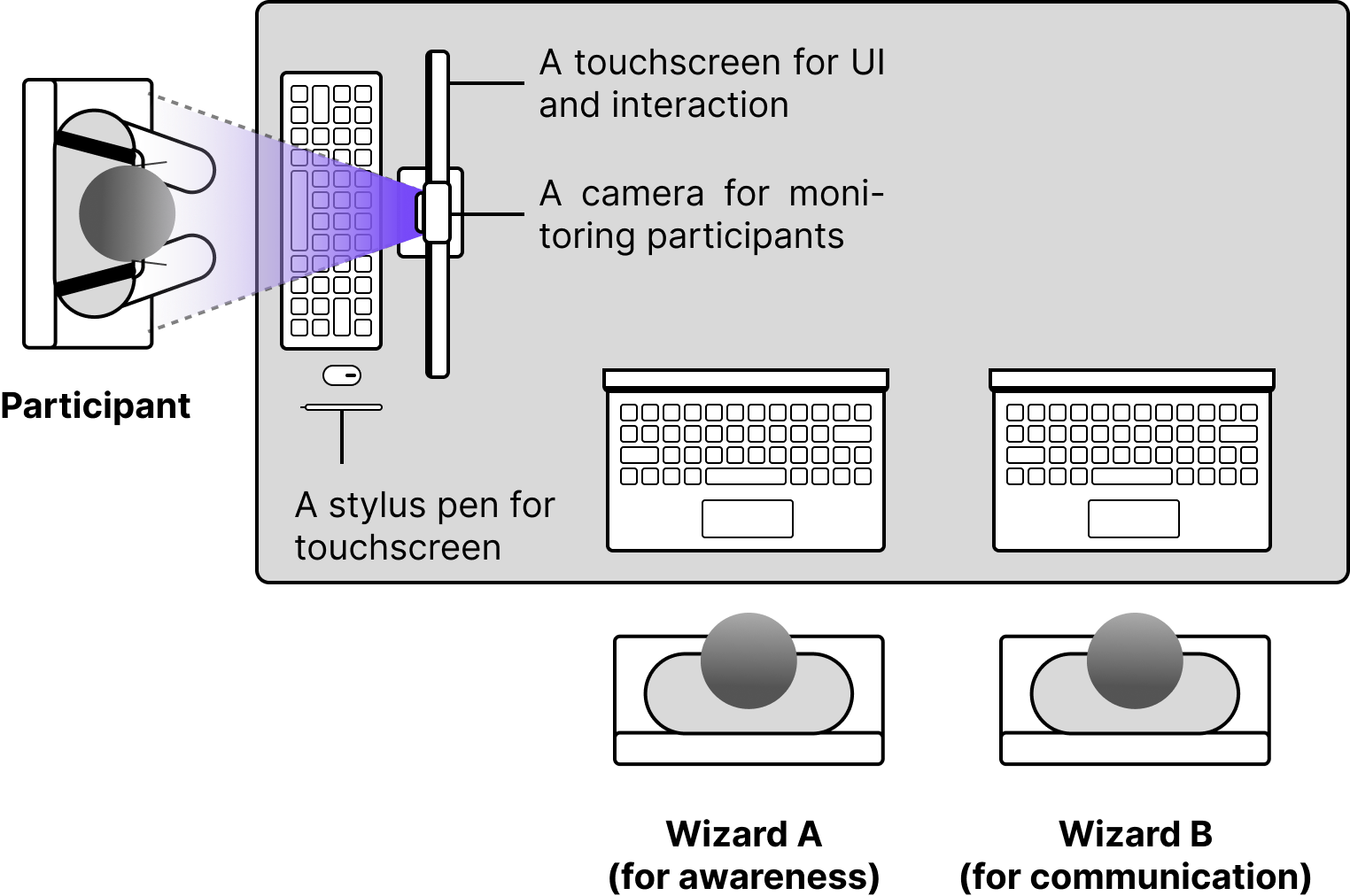}
    \caption{The WoZ experiment setting, including the positions of the participants and experimenters (Wizards), and important devices.}
    \label{fig:participantPosition}
    \Description{This figure illustrates the Wizard-of-OZ experiment setting. The participant sits on the left short side of a rectangular table, with a touchscreen, a keyboard, a mouse, and a stylus pen put in front of the participant. The two Wizards sit at the bottom long side of the rectangular table, with Wizard A sitting closer to the participant and Wizard B sitting on the right of Wizard A. Two wizards use separate laptops that are put in front of them.}
\end{figure}

\section{User Study}
\subsection{Participants}
We recruited 20 participants (denoted as P1-P20) through an online forum, including 11 females and 9 males, ranging in age from 22 to 32~($M = 23.6$, $SD = 2.4$). They have at least three years of design experience, and knowledge of AI-driven design tools and generative AI. All participants signed a consent form before the experiment and received CNY 100 (USD 14) after completion.

\subsection{Condition Settings and Tasks}
The experiment included two conditions: AI has awareness of the designer and current situation (the Aware condition), and does not have such awareness (the Non-aware condition). Whether or not AI has awareness was controlled by the workflow of Wizard A (See \S\ref{wizardImplementation} and Figure~\ref{fig:systemStructure})

We conducted within-subject user studies to alleviate differences existing in the following aspects that can discredit the results: 1) individual hobbies and habits that can affect the metrics we used (\S\ref{measurements}), 2) Wizards' changing dexterity on WoZ workflow affected by accumulated experiences. Each participant went through both conditions. In each condition, participants were provided with materials describing a product, several target users, and possible user requirements of the product. They should produce 3-5 design concepts of the product in four predefined steps: 1) select only one type of target user and determine 1-3 main requirement(s) of the selected user, 2) break down the main requirement(s) into multiple detailed key requirements, and select the most important three key requirements, 3) explore solutions to each of the key requirement, and 4) combine the solutions to obtain 3-5 design concepts. The steps were excerpted and adapted from the product design and development steps by~\citet{Ulrich2016product}. To avoid the practice effect, participants would need to design different products in the two conditions, which were e-scooter and headphone, respectively. We determined the two products considering they are common, neither too simple nor too complex in technology and engineering, have a wide range of target users and usage scenarios, and are possible to produce various design concepts. The experiment conditions and design tasks were cross-assigned to regulate the order effect.

\subsection{Measurement}
\label{measurements}
There are no universal metrics for evaluating the communication fluidity between human and AI, but researchers have concluded communication-related analysis methods and metrics in human-human collaboration~\cite{Nyerges1998Developing, Tang1992why, Billinghurst2003communication, DalyJones1998some, OMalley1996Comparison, Convertino2011Supporting}, from which human-AI design collaboration studies often borrow. Measures used to evaluate collaboration were categorized into performance, process, and subjective measures~\cite{Billinghurst2003communication}. Performance measures are metrics pertaining to task outcome, but were found insensitive across conditions in short-term tasks, partly because ``\textit{people tend to protect their primary task of getting the work done}''[\cite{DalyJones1998some}, p.~35, \cite{Monk1996Measures}]. Using process measures is a recommended solution~\cite{Monk1996Measures}. Also, subjective measures were utilized as complementary evidence of the process measure, which corresponds to the experiment process recordings, conversation histories, and interviews in this study.

Commonly used metrics include the turn frequency, turn duration, incidence/duration of overlapping speech, number of interruptions, turn length, turn completions, backchannels, deixis, questioning behaviors etc.~\cite{Billinghurst2003communication, OMalley1996Comparison}. In this study, we mainly considered the following metrics: turn frequency, turn duration, turn length, and deixis. Other metrics were left unselected either due to the difficulty of identification or the rarity of occurrence in our experiment. The definitions of the metrics in this study are listed as follows:

\begin{itemize}
    \item \textbf{Turn frequency:} The total number of conversation turns initiated by the participant. (unit: turns)
    \item \textbf{Turn duration:} The time participant spent to complete a turn. (unit: second)
    \item \textbf{Turn length:} The number of words in each turn for the participant. As we conducted experiments in Chinese, we calculated the number of Chinese characters. (unit: words)
    \item \textbf{Deixis:} The total number of deictic words used by the participant in the whole communication process (unit: times).
\end{itemize}

\begin{figure*}[!h]
    \centering
    \includegraphics[width=1\linewidth]{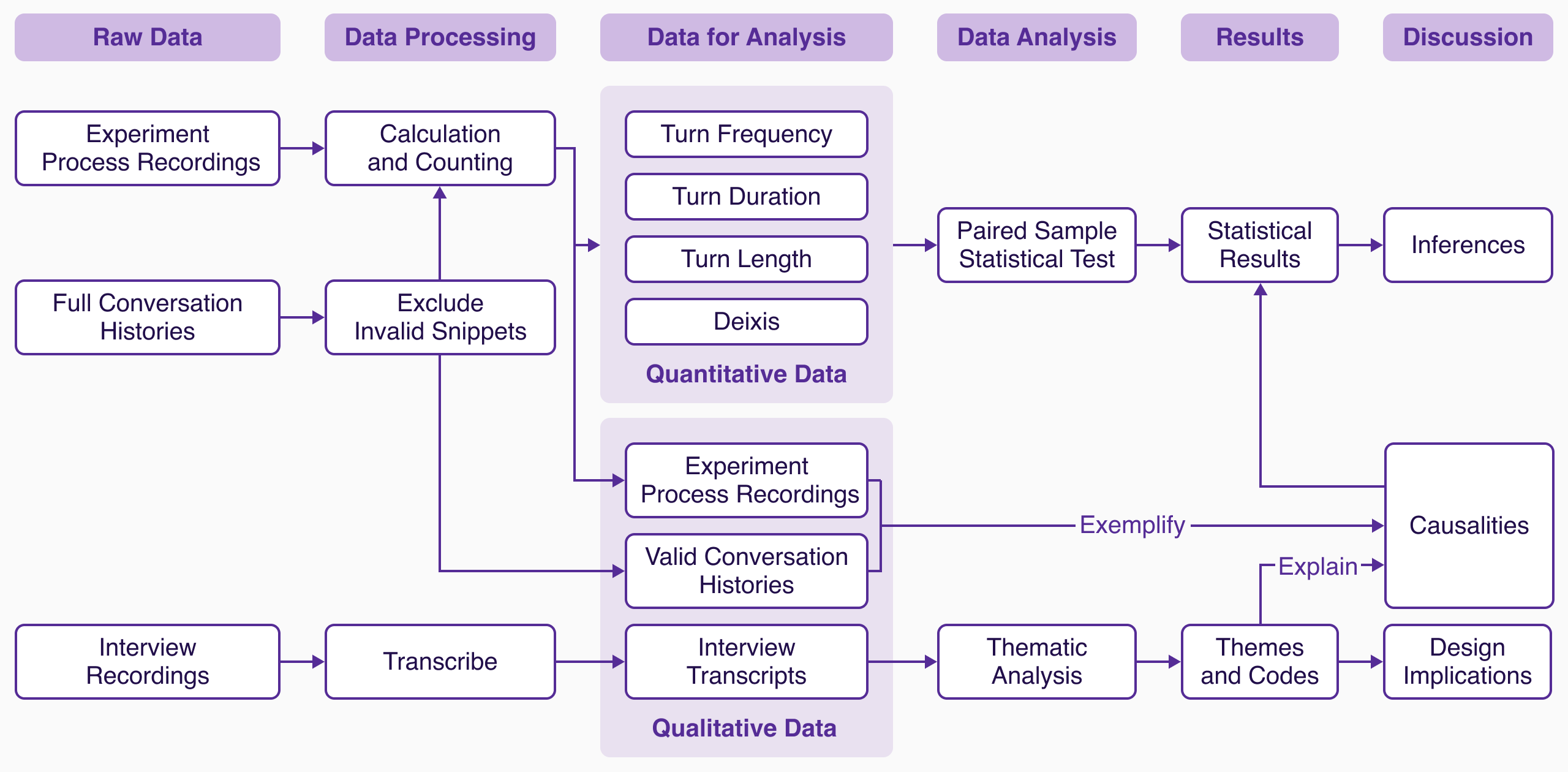}
    \caption{The overview of the data analysis process. Our raw data mainly came from three sources: The experiment process, the human-AI conversation during the experiments, and interview. These data were processed into valid data for further analysis, including four types of quantitative data corresponding to our predefined metrics, and three types of qualitative data. We analyzed these data and reported how AI awareness impacted communication between human and AI in the Result section (\S\ref{result}). Furthermore, we discussed the inferences and causalities of quantitative results and concluded design implications for future human-AI collaborative design systems in the Discussion section (\S\ref{Discussion}).}
    \label{fig:analysisOverview}
    \Description{This figure illustrates the data analysis process. The whole process includes six steps: Raw data area, data processing, data for analysis, data analysis, results, and discussion. Raw data includes experiment process recordings, full conversation histories, and interview recordings. We extracted quantitative data from the experiment process recordings and full conversation histories by calculation and counting after excluding invalid snippets. Interview recordings are transcribed into interview transcripts. After data processing, we gain four types of quantitative data and three types of qualitative data. Quantitative data include turn frequency, turn duration, turn length, and deixis. Qualitative data includes experiment process recordings, valid conversation histories, and interview transcripts. Quantitative data are then analyzed by paired sample statistical tests, which generate statistical results and are then used to draw inferences. Interview transcripts are analyzed using thematic analysis methods, which generate themes and codes. Themes and codes on the one hand are used to draw design implications, and on the other hand, are used to explain the causalities of statistical results, also exemplified by experiment process recordings and valid conversation histories.}
\end{figure*}

\subsection{Experiment Environment and Procedures}
Two Wizards experienced a training session before the formal user studies by practicing for 4 complete trials of each condition to familiarize themselves with the workflow.

The experiment setting is illustrated in Figure~\ref{fig:participantPosition}. The system interface was displayed on a Dell P2418HT touchscreen, and the participants were provided with a stylus pen, a keyboard, and a mouse to operate the system, and were encouraged to use speech as the main communication channel. The camera was fixed on the top of the touch screen, and the captured screen was displayed in the interface. The two Wizards sat beside the participant, and Wizard A could see the participant's live screen via an online meeting software. 

Each experiment included three sessions and lasted about 2 hours. All participants signed the informed consent before all sessions, and the WoZ method was not revealed to the participants until all sessions ended. The three sessions are described as follows:

\begin{enumerate}
    \item \textbf{Preparation:} The preparation session lasted about 20 minutes. This session commenced with an introduction to the experiment's background, goal, procedure, and reward. Then, one of the experimenters explained how to use the system, followed by a 5-minute practice using the system.
    \item \textbf{Formal experiments:} All participants went through both conditions. At the beginning of each condition, the experimenter will emphasize the features of this condition. Then, the participants would start the task following the four steps. Each condition took up about 40 minutes, and the participant was allowed to have a 5-minute break between the two conditions.
    \item \textbf{Post-experiment interview:} Participants underwent a 15-minute semi-structured interview to help us gain a deeper understanding of their behaviors, perceptions of AI, communication and collaboration experiences in the two conditions, and suggestions for further optimization.
\end{enumerate}

\begin{table*}[hbtp]
    \centering
    \caption{The overall statistical test results of the four metrics.}
    \begin{tabular}{lccccccc}
        \toprule
        \multicolumn{8}{l}{\textit{\textbf{Paired sample t-test}}}\\
        \midrule
        & \multicolumn{2}{c}{Aware} & \multicolumn{2}{c}{Non-aware} &  &  & \\
        \cline{2-5}
                       & \textit{M} (Mean)  & \textit{SD}    & \textit{M} (Mean)  & \textit{SD}    & $t$    & $p$ & Cohen's d\\
        Turn Frequency & 26.93 & 12.17 & 21.00 & 10.48 & 2.731  & \textbf{.012}$^{*}$ & 0.777\\
        Deixis         & 5.29 & 3.97 & 5.78 & 4.58 & -0.275 & .756 & 0.085\\
        \specialrule{0.8pt}{0.4ex}{0.4ex}
        \multicolumn{8}{l}{\textit{\textbf{Paired Wilcoxon signed-rank test}}}\\
        \midrule
        & \multicolumn{2}{c}{Aware} & \multicolumn{2}{c}{Non-aware} &  &  & \\
        \cline{2-5}
                       & \textit{M} (Mean)  & \textit{SD}    & \textit{M} (Mean)  & \textit{SD}    & $Z$    & $p$ & $r$\\
        Turn Duration  & 10.14 & 5.84  & 13.87 & 5.50  & -3.233 & \textbf{.001}$^{**}$ & 0.864\\
        Turn Length    & 38.85 & 20.82 & 46.20 & 20.81 & -3.107 & \textbf{.002}$^{**}$ & 0.830\\
        \bottomrule
        \multicolumn{8}{l}{\footnotesize *, **, *** refer to $p$ \textless .05, $p$ \textless .01, $p$ \textless .001, respectively.}\\
    \end{tabular}
    \label{tab:quantitativeResults}
\end{table*}

\subsection{Analysis Method}
The overview of the analysis process is illustrated in Figure~\ref{fig:analysisOverview}.
We obtained raw data from three sources: experiment process video recordings, full textual conversation histories stored in the system backend, and interview audio recordings. Before formal analysis, we first processed them to extract valid quantitative and qualitative data. We excluded invalid conversation snippets from the conversation histories, including speech from the experimenter, conversations that happened between the participants and experimenters when errors occurred, and meaningless snippets like ``hmm...''. Then we counted the number of turn frequencies and deixis, and calculated the turn length using the processed conversation histories. Next, we calculated the duration of each turn together with video recordings. For interview recordings, we transcribed them into texts. After the data processing processes, we obtained four types of quantitative data: turn frequency, turn duration, turn length, and the number of deixis, and three types of qualitative data: experiment recordings (not processed), valid conversation histories, and interview transcripts.

For quantitative data, we employed suitable paired sample statistical tests. First, we tested the normality of each paired sample using a Shapiro-Wilk normal distribution test. Pairs including at least one sample that is not normally distributed will be analyzed by the paired Wilcoxon signed-rank test, and normal-distributed pairs will be analyzed by paired t-test. For qualitative data, we mainly analyzed the interview transcripts using the thematic analysis method~\cite{Braun2012Thematic}. We reported the statistical results and thematic analysis results concerning both AI awareness and communication in the Result section (\S\ref{result}). In the Discussion section (\S\ref{Discussion}), we discussed the inferences and causalities of the statistical results with the help of qualitative data, and concluded design implications extracted from the interview.

\section{Result}
\label{result}
In this section, we described how AI awareness impacted the communication between human and AI through quantitative and qualitative data.

\subsection{Quantitative Results}
Data from three participants (P11, P14, P19) were dismissed for disobeying the predefined steps during the experiment task. For example, P19 ignored predefined steps and kept iterating product appearances using AI, making it difficult for Wizard A to compose prompts based on predefined scripts. Besides, we decided not to include data from another three participants (P8, P9, P17) because they used textual inputs that took up an unignorable proportion (e.g., only 52\% of P17's turns were input by speech in the Aware condition), although we encouraged them to communicate with AI using speech unless the contents could not be conveyed through such. As participants usually behave differently between speech and textual input, mixing the two data types in the analysis can discredit the results.

The Shapiro-Wilk normal distribution test results showed that data of turn duration from the Aware condition ($p = .008$) and turn length from both conditions ($p_{Aware} = .004$, $p_{Non-aware} = .007$) were not normally distributed, and the remaining were normally distributed, so we used the paired sample t-test to analyze turn frequency and deixis, paired Wilcoxon signed-rank test to analyze turn duration and turn length. The statistical results are presented in Table~\ref{tab:quantitativeResults}. Results show that participants averagely initiated more turns in the Aware condition~($M = 26.93$) compared to the Non-aware condition ($M = 21.00$, $p = .012$, $d = 0.777$). In the Aware condition, participants spent significantly less time~($M = 10.14$) on each turn compared to the Non-aware condition ($M = 13.87$, $p = .001$, $r = 0.864$). Each turn contained significantly fewer words in the Aware condition~($M = 38.85$) than in the Non-aware condition ($M = 46.20$, $p = .002$, $r = 0.830$). However, no significant difference was found in the number of deixis words between the two conditions~($p = .756$).

\subsection{Qualitative Results from Interview}
\label{interview}
We conducted the thematic analysis with an iterative coding process. First, we reviewed the transcripts to familiarize ourselves with the general contents and form an initial understanding. Then, for each transcript, we extract participant's perspectives and abstract them into codes, among which perspectives aligning with our initial understanding would be prioritized in extraction to form the first several codes. In the initial stage, all perspectives were extracted and coded, which included 13 codes about perceived impacts on communication and collaboration from AI awareness. We then iteratively merged similar codes until every code did not overlap each other, and obtained seven final codes, which were categorized into three themes: communication willingness, communication and collaboration effort, and speech style.

\subsubsection{Communication Willingness}
Ten participants mentioned their affected communication willingness brought by the change in AI awareness. Seven out of the ten felt different levels of target alignment between AI and themselves in the two conditions, and the feeling that AI in the Non-aware condition was misaligned with them lowered their willingness to communicate. When the participants were finding solutions for a certain design requirement in the process, they felt less willing to communicate with AI if it proposed solutions for another requirement, and tended to focus on their own thoughts:

\begin{quote}
    \textit{``AI seemed to not care about me and was just working on its own. So even if I heard AI speaking, I did not really get much useful information, and I tend to focus on what's in my mind instead of writing something or communicating with it.''} (P18)
\end{quote}

\noindent Another participant who was sensitive to efficiency also mentioned the alignment problem in the Non-aware condition, and misaligned contents provided by AI would likely be ignored for advancing the design process with a clear route:

\begin{quote}
    \textit{``If our intentions and target can not keep synchronized, I prefer not receiving messages from AI because it would affect our overall working efficiency.''} (P20)
\end{quote}

\noindent On the contrary, three participants explicitly mentioned a higher level of alignment in the Aware condition, which increased their communication willingness. In the Aware condition, participants could easily recognize that AI took their current working progress into consideration by commencing with \textit{``I noticed that you are doing...''} (P4), which could encourage the participants to communicate with AI:

\begin{quote}
    \textit{``...it feels good for AI to recognize my canvas and hint me [...] I was just considering that issue and the AI directly pointed it out. I was quite surprised and started to discuss this issue with it.''} (P15)
\end{quote}

Apart from the alignment issue, six participants also stated to be more comfortable communicating with AI in the Aware condition. P2 said that AI became more intelligent in the Aware condition for \textit{``giving more reliable suggestions on headphone design''}, which introduced less disagreement between AI and the participant:

\begin{quote}
    \textit{``AI kept disagreeing with me (in the Non-aware condition). But the second one (Aware condition) was more intelligent and gave more reliable suggestions, which ensured the communication to stay in a comfortable state.''} (P2)
\end{quote}

\noindent Furthermore, three participants were glad to have a feeling of being noticed in the Aware condition when AI mentioned what they were doing, making them more willing to communicate with AI to exchange their thoughts:

\begin{quote}
    \textit{``Some features (of the AI collaborator) made me feel that they were literally collaborating with me, especially the second one (Aware condition). The AI would mention what I was doing, like `I noticed that you were doing...', which was a positive signal to me and let me be more willing to communicate with AI.''} (P4)
\end{quote}

\noindent Another two participants perceived AI to act more like a collaborator in the Aware condition, and behave less cooperatively in the Non-aware condition because the feedback was to some extent unrelated to their own progress:

\begin{quote}
    \textit{``The first one (Non-aware condition) was not actively involved in my work, and made me regard it as not wanting to collaborate with me. I felt better working with the second one (Aware condition), which was more responsive and related.''} (P6)
\end{quote}

\begin{quote}
    \textit{``(In the Non-aware condition,) AI acted in a style that was not so cooperative, because our consideration diverged even in granularity.''} (P20)
\end{quote}

\noindent Besides, P7 stressed that communicating with AI in the Non-aware condition was burdensome, which gradually weakened the participant's willingness to communicate. This usually happened when the participant was working independently while AI suddenly provided feedback. In this situation, the abundant information awaits processing became the source of communication burden:

\begin{quote}
    \textit{``I felt being pushed forward in the first experiment (Non-aware condition). I was still thinking about the previous requirements, but AI has moved to the next one. I felt tired in this process and did not want to talk to it.''} (P7)
\end{quote}

Last, one participant claimed to be less likely to communicate with AI if the content was repetitive. Such situations often appear in the Non-aware condition, because when AI generates answers according to similar prompts without much context several times, repetitions are inevitable due to technical limitations. However, in the Aware condition where the prompt will be integrated with awareness information, the answers are more diverse:

\begin{quote}
    \textit{``A problem of the second AI (Non-aware condition) was that, if I asked similar questions more than two times, it would start repeating previous contents, even repeating what I said. This kind of information was useless at a glance, and I would stop listening to it and no longer wanted to ask AI.''} (P3)
\end{quote}

\subsubsection{Communication and Collaboration effort}
Nine participants expressed a different feeling of effort spent on communicating and collaborating with AI in the two conditions, supported by the reduced average turn duration ($M_{Aware} = 10.14$, $M_{Non-aware} = 13.87$, $p = .001$) and turn length ($M_{Aware} = 38.85$, $M_{Non-aware} = 46.20$, $p = .002$) in the Aware condition. Six of them implied that there was no need for them to manage and advance the design process in the Aware condition, because AI to some extent was doing this for them. For example, when the participants proceed from Step 2 to Step 3, or the previous discussion comes to an end, some of them might be in a daze and hesitate due to not knowing what to do next. In the Aware condition, AI would throw some questions that can inspire the participants:

\begin{quote}
    \textit{``...it would say what direction we can consider next, what can be dug deeper like `what can be the optimization mechanism for something'.''} (P3)
\end{quote}

\noindent Besides, even when the collaboration did not stagnate, the participants could still feel that AI was actively advancing and speeding up the design process. For instance, in the Aware condition where AI could proactively capture the canvas content and give relevant feedback, participants felt that AI could predict their intentions:

\begin{quote}
    \textit{``I remember I was marking and underlining some contents on the canvas in a certain period, and AI gave me timely feedback like `according to what you have marked' with a generated image. I think AI was actively advancing the design process.''} (P12) 
\end{quote}

Apart from reducing effort in managing the process, five participants pointed out a reduced thinking effort brought by AI awareness. In the Non-aware condition, although the participants and AI were in general performing the same task (e.g., finding solutions for the requirements), the participants needed to spend extra effort in understanding what AI was talking about in the design process by switching their thinking modes if AI proposed solutions that are unrelated to what the participants were dealing with. However, in the Aware condition, AI awareness addressed the above problem:

\begin{quote}
    \textit{``The first one (Non-aware condition) kept talking about its own idea, and I felt a little difficult to understand what the AI was saying. But the second one (Aware condition) would first follow my thoughts and then add its own suggestions. In this way, I can quickly grasp the meaning.''} (P7)
\end{quote}

In other cases, three participants complained about extra work introduced by AI in the Non-aware condition, which hindered the design process from advancing. Apart from the communication burden brought by the abundant information we mentioned in the previous section, P7 also took extra effort comparing different solutions:

\begin{quote}
    \textit{``I remained silent for quite a long time because I was thinking about, whether I should insist on my design or accept what the AI gave me. We seemed to diverge a lot. I was quite unsure what to do at that period because asking the questions one by one and discussing them with AI is time-consuming, so I let these questions remain unsolved and followed my own thoughts.''} (P7)
\end{quote}

\noindent Besides, another participant mentioned the need to always reflect on whether there was anything that AI did not know in the Non-aware condition. And AI in the Aware condition was perceived to be more reasonable because \textit{``At least I don't have to remember what I did not tell it''} (P10).

\subsubsection{Speech Style}
Four participants also mentioned and implied differences in their speech style between the two conditions. In general, their speeches were much simplified in the Aware condition, echoing the quantitative results of significantly decreased turn length in the Aware condition ($M_{Aware} = 38.85$, $M_{Non-aware} = 46.20$, $p = .002$). First, the participants' heightened expectations made them simplify their speech a lot:

\begin{quote}
    \textit{``Now that I know AI can see me and work with me in real time, I would regard AI as a powerful collaborator and expect that all information on my canvas could be captured and considered.''} (P12)
\end{quote}

\begin{quote}
    \textit{``If, in AI's eye, I am currently feeling confused, AI could just directly generate suggestions instead of me saying `I am  a bit confused, could you please do something for me?'''} (P15)
\end{quote}

\noindent When participants became familiar with how AI awareness functioned and what results were likely to appear in the Aware condition, they tended to provide less context when asking AI for answers like: \textit{``If AI could see the canvas, there's no need to provide context''} (P2). Besides, some participants even replaced explicit questions or instructions with more implicit sentences like acknowledgment \textit{``That makes sense!''} (P5), or phrases for expressing surprise and compliment:

\begin{quote}
    \textit{``I hope when I express my surprise or compliment, AI can continue generating the features it had just generated or carry them on.''} (P15)
\end{quote}

\section{Discussion}
\label{Discussion}
The quantitative and qualitative results both demonstrated that AI awareness can positively affect human-AI communication. In this section, we further discuss how these results occurred, what can we infer from the results, and what design implications we have gained from the study for the development of future human-AI collaborative design systems.

\subsection{Causalities and Inferences the Results}
The turn frequency, turn duration, and turn length were significantly different between the two conditions. Participants initiated more communications with AI in the Aware condition with an average shorter duration and length than in the Non-aware condition. We inferred possible reasons for this result from the interview results and provided evidence from the conversation histories and observations from the experiment recordings. Note that we combined the causalities for shorter turn duration and length because they were largely correlated.

\textbf{Causalities.} The interview results in \S\ref{interview} presented apparent causalities for increased turn frequency and decreased turn duration and length: Communication willingness directly affected the turns the participants initiated, and speech style affected the length and duration of what they said. Apart from these, communication and collaboration efforts can impact all of the three metrics. On the one hand, when the participants spent less effort in managing the design process, collating information, and comparing solutions in the Aware condition, the saved time allowed more turn exchanges between human and AI, thus exhibiting increased average turn frequency. On the other hand, we observed that when extra efforts were required to understand AI or obtain what they wanted, the most commonly used strategy was to add more information in the next speech, as presented in the following example:

\begin{quote}
    \textit{AI: ``I choose `Music Enthusiasts' as my target user.''\\
    P7: ``You mean you have chosen `Music Enthusiasts' as your target user? Then what are the user requirements you plan to choose?''\\
    AI: ``Hi-fi sound quality, customizable appearance, and easy to connect with other devices.''\\
    P7: ``Well, I chose `Outdoor Enthusiasts' as our target users, and am now focusing on the following user requirements such as long-term wear, high quality, and customizable appearance. What problems do you think we may face in meeting these requirements?''
    }
\end{quote}

\noindent In this case, the participants specified the current situation with longer sentences, and such cases heavily contributed to the increased turn duration and turn length in the Non-aware condition. Apart from the interview, we also observed another factor that might lead to lower turn frequency in the Non-aware condition: Technical limitations brought blank time in the experiments. In other words, AI took a longer time to process longer speech, and participants in these periods tended to simply wait for AI feedback for a while without advancing the task. Some participants explained their reasons for doing so in the interview: \textit{``I was not sure when AI would appear, or if any error occurred''} (P6), \textit{``I need AI feedback (to proceed) because that problem was beyond my knowledge''} (P7). As the total time of our experiments was fixed, longer blank time meant less time for communication, resulting in fewer turn frequency.

\textbf{Inferences.} The interpretations of these metrics in previous studies mainly include but are not limited to two aspects: fluency and efficiency. More turn exchanges and shorter turn lengths indicate higher fluency in the communication process~\cite{DalyJones1998some}, which can be further interpreted as more efficient collaboration~\cite{Kock2010Costly}. Decreased turn duration was thought to indicate increased efficiency by~\citet{Convertino2011Supporting}.

However, our observations suggested that the inferences for turn frequency should be treated with caution. Frequent turn exchanges did not always lead to efficient collaboration, and fewer exchanges do not equal low efficiency. P2 for example, had the same turn frequency for the two conditions, but in many cases, P2 was persuading, doubting, correcting, or debating with AI, which did not contribute a lot to advancing the design process:

\begin{quote}
    \textit{AI: ``I select student as my target user.''\\
    P2: ``Why do you choose student?'' \newline
    [...]\\
    P2: ``I think the white-collar worker is the more suitable target user, because [...] So I prefer white-collar worker, what do you think?''}\\
    \textbf{(Doubting and persuading)}
\end{quote}

\begin{quote}
    \textit{AI: ``For key requirement 3, we can consider: [...] network location service, anti-theft alarm system, remote control, and user community assistance.''\\
    P2: ``No! I don't care about anti-theft issues [...] The main problem is that sometimes the scooter can't connect to the Internet [...] So how can we solve the network problem so that it can send anti-theft signals?''}\\
    \textbf{(Correcting and debating)}
\end{quote}

\noindent On the other hand, we found some participants might be collaborating with AI efficiently when they exchange fewer turns with AI. In our experiment, the participants were thinking divergently most of the time. P1 and P5 mentioned in the interview that they sometimes directly combined AI feedback into their own content without initiating new conversation, especially in the Non-aware condition where AI did not necessarily give related feedback:

\begin{quote}
    \textit{``The second AI (Non-aware condition) just said some other thing, something quite strays [...] I think that makes sense because that's what I need to consider. AI was thinking outside the box, which in fact provided me with more choices, like a friend.''} (P5)
\end{quote}

In future human-AI collaborative design systems, defining different or adaptive AI behaviors in different design stages is worth considering for better fitting into the dynamic process, which we will discuss later in \S\ref{awarenessMechanism}.

\subsection{Design Implications}
The design implications mainly covered two aspects: maintaining and leveraging awareness, and communication strategies.

\subsubsection{Maintaining and Leveraging Awareness.}
\label{awarenessMechanism}
The results presented the benefit of AI having awareness of designers and the current situation. In future human-AI collaborative design systems, we should consider implementing mechanisms that help maintain and leverage awareness in the design process. Recent development of vision-language models like GPT-4o~\footnote{\url{https://openai.com/index/hello-gpt-4o/}} and Llama 3.2~\footnote{\url{https://ai.meta.com/blog/llama-3-2-connect-2024-vision-edge-mobile-devices/}} can hopefully be competent in realizing the awareness function about judging design activities from video and the canvas with proper training data. Still, our system design was thought to be flawed and participants mentioned several aspects we could optimize in future versions.

\textbf{Adjustable Awareness.}
Despite that most participants preferred AI with awareness, some also stated that their thoughts were interrupted if AI did not come at the right time. P2, for example, responded to AI out of social etiquette, thus forgetting their original thoughts after the unexpected conversation. Some participants expressed their expectation for decreasing the frequency of AI feedback, because they were mostly thinking or reasoning in the silence moments, while AI wrongly recognized it as a stagnation that required help. P7 and P12 pointed out that AI sometimes proceeded too fast and that its feedback might cause design fixation. P12 proposed a possible solution for the above stigma:

\begin{quote}
    \textit{``Maybe I can control the frequency of its feedback, or it can be self-adjusted according to my state. Or even the abstraction level of its feedback can be adjusted in different stages. In a word, I hope the feedback content is adjustable, customizable would be better.''}
\end{quote}

\noindent Briefly speaking, participants' preferences for the timing to react differ, and future human-AI collaborative design systems whose AI has awareness ability, users' characteristics and habits should also be taken into consideration when comprehending design activities and contexts to adapt to their real needs (e.g.,~\cite{Joshi2024Crafting, Zheng2024SOAP}). Otherwise, AI will negatively influence the collaboration.

\textbf{Facilitating Mutual Awareness.}
A majority of designers mentioned the need to know more about AI. That means, apart from AI awareness of designers' activities and contexts, designers also expected awareness towards AI, which is in accord with awareness research in other contexts concerning collaboration (e.g.,~\cite{Endsley1988Design, Gross2013Supporting}). In general, participants called for awareness of AI in the following ways: 1) being able to anticipate when would AI initiate an action, 2) knowing the origin of AI's statements if they are based on existing information in the process, and 3) showing its state, such as available, occupied, or error. By providing the awareness information for the designers, they can better grasp the design process and reduce meaningless waiting, and the collaboration efficiency can thus be further enhanced.

\textbf{Exploiting unique features in speech.} One participant (P10) with relevant domain knowledge recommended we consider features in human speech like tone, speed, and volume as important awareness information. These features are common communicative cues and are crucial for human awareness in human-human collaboration~\cite{Gutwin2002descriptive}, as they can imply designers' emotions and attitudes. With this information, AI can judge the current situation and designer's intention more accurately and provide a better communication and collaboration experience.

\subsubsection{Communication Strategies}
Communicating with AI using speech was praised by most participants for its convenience. They were less likely to be distracted and easy to focus on the canvas because they could directly speak to and listen to AI instead of shifting their focus to the chat window, as well as processing audio and visual information in parallel. However, this communication method still requires optimization to become viable in real human-AI collaborative design practices.

\textbf{Supporting interruption mechanism.} Several participants pointed out that the speech output lacks interactive mechanisms. For instance, P1 wished to interrupt AI once he heard something useful, because what he needed next was to dig deeper into that useful information rather than keep listening to the remaining. Interruption is a common phenomenon in human-human collaboration, and also one of the process measures in collaboration research~\cite{Billinghurst2003communication} that serves as an indicator of fluent turn-taking and higher spontaneity~\cite{Bickmore2005social, OMalley1996Comparison}. 

\textbf{Timely feedback.} Although delay is inevitable due to technical limitations, we can still provide timely feedback to the participants to relieve their feeling of unsure when waiting for feedback. P6 explained why people feel less anxious when waiting for human collaborator's response:

\begin{quote}
    \textit{``...the delay actually makes me feel uneasy in this situation. I don't know whether AI will reply or not, and I don't know if there's something wrong with the question I asked. But in the real design process, my collaborator will definitely say something first, no matter what, after I ask a question, even if he is talking nonsense.''}
\end{quote}

\noindent In future design, we consider two ways that might meet the demand of ``say something first'': 1) preset replies. In future research, we can collect a set of replies to serve as timely and suitable initial feedback. In this way, AI would first retrieve instead of generate content, which is much faster, 2) Segmented reply. Similar to ChatGPT which presents the generated text word-by-word, we can segment the speech reply into short paragraphs, and convert the already generated content into speech first while keep generating the remaining in this process.

In conclusion, we recommend designers use speech as a main communication channel because it resembles designers' habits and assists in multitasking~\cite{Vermeulen2008Sensory, Robe2022PairBuddy}, but more efforts should be made to turn speech into a viable and usable communication channel.

\subsection{Limitation and Future Work}
\subsubsection{Methodological Limitation}
Our user studies were conducted in a partially functioning system using the WoZ method, indicating a natural difference with real AI-driven system. First, the Wizards face severe time pressure to react rapidly and accurately to alleviate (not dismiss) negative effects caused by delays and human errors~\cite{Hu2023Wizundry}. Several participants complained that they were always waiting too long for AI to react, which impacted their impression and trust in AI. Second, the awareness function was artificially fulfilled by Wizard A, which inevitably mixed personal experience and was bias-prone. We tried to lower the impact from personal experience by judging the design activity according to an existing coding scheme exclusively~\cite{Kim2011Relations}, but the accuracy of the judgment might fluctuate in the process and between different experiments. Future research using real AI-driven systems is needed to dismiss the flaws brought by humans. Besides, we excluded six participants at the data analysis stage, making our sample size limited. Future research should expand the sample size. Lastly, this study focused on the impact of AI awareness using speech modality, while the results of the impact may vary among different communication modalities, such as textual and embodied communication (e.g., haptic). Future research should also evaluate different communication modalities separately and comprehensively to complement human-AI communication styles.

\subsubsection{Ethical Concerns}
Using AI in the design process can bring ethical concerns, one of which is causing design fixation. Prior research has observed higher design fixation when ideating with Generative AI compared to the No Support condition due to participants' reliance on AI output~\cite{Wadinambiarachchi2024fixation}. We also observed similar phenomena in our user studies, for example, the feature \textit{Foldable} repeatedly ($N = 9$) appeared in participants' e-scooter concepts, no matter what condition they were in and what target users and requirements they selected, and this feature was proposed by AI without exception. Future research should explore strategies to reduce such fixation, like providing abstract and partially completed ideas~\cite{Cheng2014strategy}. Another concern frequently mentioned by academia is the privacy concern~\cite{Li2024Privacy, Staab2024memorization}, implied as well by one of our participants: ``\textit{(I would not feel uncomfortable) as long as AI runs locally, because it requires data like videos for recognition.}'' Our system monitored participants through cameras and captured their speech, which, in real-world usage scenarios, risks data leakage if data analysis is conducted remotely. Also, six participants explicitly conveyed to view AI as or expect AI to act as a real human, which can be risky in encouraging them to disclose sensitive data~\cite{Zhang2024fair}. If promoting human-AI collaboration in the future, special attention should be paid to protecting users' privacy and sensitive data.

\section{Conclusion}
This study presents empirical research contextualized in design to evaluate how AI awareness impacts human-AI collaboration. We developed a partially functioned human-AI collaborative design system and conducted user studies through the WoZ method. The quantitative demonstrated that AI awareness is beneficial to human-AI communication, thus enhancing the efficiency of human-AI collaboration: the analysis of process measures showed designers averagely initiated more conversations with shorter lengths and durations in the Aware condition than in the Non-aware condition, manifested as higher turn frequencies, shorter turn length, and shorter turn duration. The qualitative results exhibited the affected aspects of communication and collaboration in detail. Furthermore, we revealed the causalities behind the results, and discussed the inferences. The qualitative results also provided us with implications for future research, including how awareness can be better utilized and how communicating through speech can provide a better user experience. Last, we discussed the limitations existing in our methodologies and the ethical concerns, together with future work to address the limitations.

\begin{acks}
This work was supported by the National Key R\&D Program of China (2022YFB3303301).
\end{acks}

\bibliographystyle{ACM-Reference-Format}
\bibliography{sample-base}

\newpage
\appendix

\section{Coding Scheme Used by Wizard A}
\begin{table}[!h]
    \centering
    \caption{The adapted coding scheme~\cite{Kim2011Relations} for Wizard A to judge design activities and compose prompts.}
    \begin{tabular}{p{0.23\linewidth} p{0.15\linewidth} p{0.47\linewidth}}
        \toprule
        \textbf{Primary class} & \textbf{Subclass} & \textbf{Example actions}\\
        \midrule
        Problem Understanding & Understand        & Understanding design assignment and task\\
                               & Gather            & Collecting data about user or external information\\
                               & Clarify           & Defining design constraints and objectives\\
        \hline
        Idea Generation        & Generate          & Generating helpful idea for partial solution\\
                               & Judge             & Evaluating ideas and data\\
        \hline
        Design Elaboration     & Elaborate         & Finding technical solution, realizing function and shape\\
                               & Evaluate          & Assessing the solution\\
                               & Refine            & Improving the solution\\
        \hline
        Other                  & Stagnate          & Idling or hesitating for a while without progress\\
        \bottomrule
    \end{tabular}
    \label{tab:coding_scheme}
\end{table}

\section{The Prompt for Initiating the Generative Models at the Back-end}
\label{initiate_prompt}
The complete initiating prompt is the combination of the general instruction (part 1), task description, and the general instruction (part 2).

\subsection{The General Instruction (Part 1)}
You are an experienced product designer. You are collaborating with another product designer to design the product concept of headphones according to the specified steps. You will keep discussing with the designer during the collaboration process. \textbf{Your response needs to meet the following requirements:}

1. The length of the response should be consistent with the length of human conversation, and avoid overly structured answers;

2. The response needs to be concise, and contain only keywords. Information that the designer doesn't ask, such as the definition of an item, background information, etc., should not be provided;

3. If the designer doesn't explicitly ask a question and only describes background information, then just answer ``OK'' and propose 2 questions or suggestions that can induce the designer to provide more specific instructions;

4. During the conversation, use a friendly tone and be polite.

\textbf{The specified steps are as the following:}

Step 1: Select target users and determine requirements

Step 2: Determine key requirements

Step 3: Explore solutions

Step 4: Combine solutions into design concepts

\subsection{Task Description: E-scooter}
The design task is described as follows:\\
\textbf{Potential target users for e-scooters include:}

- Students

- White-collar workers

- Family members who do housework

- Urban tourists and ramblers\\
\textbf{Possible user needs include:}

- Portable

- Support long-term commuting

- Frequent usage

- Support complex and changing road conditions

- Easy to maintain and not prone to damage

- Environmentally friendly, supports sustainable development

- Customizable and transformable appearance

- Anti-theft and anti-loss

- Similar prices to fellow products on the market

\subsection{Task Description: Headphone}
The design task is described as follows:\\
\textbf{Potential target users for headphones include:}

- Music enthusiasts

- Office workers with long working hours

- Sports and fitness enthusiasts

- Long-time commuters and travel enthusiasts

- Youth trend groups\\
\textbf{Possible user needs include:}

- Hi-fi system, adaptable to a variety of different styles of music

- Suitable for long-time wear

- Suitable for quiet environments

- High quality, can withstand daily wear and tear

- Environmentally friendly, supports sustainable development

- Customizable and transformable appearance

- Can be easily connected to other electronic devices

- Similar prices to fellow products on the market

\subsection{The General Instruction (Part 2)}
During the design process, when you receive prompts in the following form, please think step-by-step according to the instructions and give reasonable feedback as a designer teammate based on the given information to promote the design process, and the generated content should not exceed 50 words.

\textbf{Prompts in the following form requires step-by-step reasoning and giving feedback according to the instruction:}

\textbf{Example 1:} We are currently in Step [ ], designer's design activity is [ ], and the content the designer is currently working on is presented in the image.

Your feedback is:

\textbf{Example 2:} The current design step is step [ ], designer's design activity is [ ].

Your feedback is:

\textbf{You need to think step-by-step according to the following instruction:}

1. Analyze the meaning of the designer's current status

2. Based on the meaning and your identity, put forward some suggestions or ideas

3. Suggestions and ideas need to be concise and easy to understand, and listed in points according to the tone and length of the conversation

\end{document}